\documentclass[
12pt,tightenlines,%
preprintnumbers,%
showpacs,%
prl]{revtex4}
\usepackage{graphicx}
\newcommand{\be}{\begin{equation}}
\newcommand{\ee}{\end{equation}}
\newcommand{\bea}{\begin{eqnarray}}
\newcommand{\eea}{\end{eqnarray}}

\newcommand{\0}{\over }

\newcommand{\g}{g_{\rm eff}}
\newcommand{\geff}{g_{\rm eff}}

\newcommand{\im}{{\rm Im}}
\newcommand{\re}{{\rm Re}}

\begin{document}
\preprint{TUW-03-23}
\title{Anomalous specific heat in 
high-density QED and QCD}
\author{A. Ipp}
\affiliation{Institut f\"ur Theoretische Physik, Technische
Universit\"at Wien, \\Wiedner Haupstr.~8-10, 
A-1040 Vienna, Austria }
\author{A. Gerhold}
\affiliation{Institut f\"ur Theoretische Physik, Technische
Universit\"at Wien, \\Wiedner Haupstr.~8-10, 
A-1040 Vienna, Austria }
\author{A. Rebhan}
\affiliation{Institut f\"ur Theoretische Physik, Technische
Universit\"at Wien, \\Wiedner Haupstr.~8-10, 
A-1040 Vienna, Austria }
\begin{abstract}
Long-range quasi-static 
gauge-boson interactions
lead to anomalous (non-Fermi-liquid)
behavior of the specific heat in the
low-temperature limit of an electron or quark gas with
a leading $T\ln T^{-1}$ term. We obtain perturbative
results beyond the leading log approximation and find that
dynamical screening gives rise to 
a low-temperature series involving also anomalous
fractional powers $T^{(3+2n)/3}$.
We determine their
coefficients in perturbation theory up to and including order $T^{7/3}$ and
compare with exact numerical results obtained in the large-$N_f$ limit
of QED and QCD.
\end{abstract}
\maketitle

It has been established long ago \cite{Holstein:1973} in the
context of a nonrelativistic electron gas that the only
weakly screened low-frequency transverse gauge-boson interactions
lead to a qualitative deviation from Fermi liquid behavior.
A particular consequence of this
is the appearance of an anomalous contribution to the 
low-temperature limit of entropy and specific
heat proportional to $\alpha T\ln T^{-1}$
\cite{Holstein:1973,Gan:1993,Chakravarty:1995}, but it was argued that
the effect would be probably too small for experimental detection.

More recently, it has been realized that analogous
non-Fermi-liquid behavior in ultradegenerate QCD
is of central importance to the magnitude of the gap in
color superconductivity
\cite{Son:1998uk,Brown:1999aq,
Wang:2001aq},
and it has been pointed out \cite{Boyanovsky:2000bc} 
that the anomalous contributions
to the low-temperature specific heat may be of interest in
astrophysical systems such as neutron or protoneutron stars,
if they involve a normal (non-superconducting) 
degenerate quark matter component.

So far only the coefficient of the $\alpha T\ln T^{-1}$ term
in the specific heat
has been determined (with Ref.~\cite{Chakravarty:1995} correcting
the result of Ref.~\cite{Holstein:1973} by a factor of 4), but not
the complete argument of the leading logarithm.
While the existence of the $T\ln T^{-1}$ term implies that there
is a temperature range where the entropy or the specific heat
{\em exceeds} the ideal-gas value, without knowledge
of the constants ``under the log'' it is impossible
to give numerical values for the required temperatures.

Furthermore, a quantitative understanding of
these anomalous contributions is also of interest
with regard to the recent progress made in high-order
perturbative calculations of the pressure (free energy) of
QCD at nonzero temperature and chemical potential \cite{Vuorinen:2003fs},
where it has been found that dimensional reduction techniques work
remarkably well except for a narrow strip in the $T$-$\mu$-plane
around the $T=0$ line.

In the present Letter we report the results of a calculation
of the low-temperature entropy and specific heat for
ultradegenerate QED and QCD
which goes beyond the leading log approximation. Besides completing
the leading logarithm, we find that for $T/\mu \ll g \ll 1$, where
$g$ is either the strong or the electromagnetic coupling constant,
the higher terms of the low-temperature series involve
also anomalous fractional powers $T^{(3+2n)/3}$, and we give
their coefficients through order $T^{7/3}$.

Our starting point is an expression for the thermodynamic potential
of QED and QCD which becomes exact in the limit of large flavor number $N_f$
\cite{Moore:2002md,
Ipp:2003jy},
and which  at finite $N_f$ 
has an error of order $g^4$ in the regime $T/\mu \ll g$,
\begin{eqnarray}
{P}  &=& NN_f\left( {\mu^4\012\pi^2}+{\mu^2 T^2\06}+ {7\pi^2 T^4\0180}\right)
\nonumber\\&&
 -N_g\int \frac{d^{3}q}{(2\pi )^{3}}\int _{0}^{\infty }\frac{dq_{0}}{\pi } \nonumber \\
&\times&\!
\biggl[2\left( [n_{b}+\frac{1}{2}]\textrm{Im}\ln D^{-1}_T
-\frac{1}{2}\textrm{Im}\ln D^{-1}_{\rm vac}\right)  \nonumber\\
 & +&\!\! \left( [n_{b}+\frac{1}{2}]\textrm{Im}\ln 
\frac{D^{-1}_L}{q^{2}-q_{0}^{2}}
-\frac{1}{2}\textrm{Im}\ln \frac{D^{-1}_{\rm vac}}{q^{2}-q_{0}^{2}}\right) \biggr] \nonumber\\
&&+O(g^4 \mu^4),\qquad (T/\mu \ll g)
\label{pressurecomplete}
\end{eqnarray}
where $N=3$, $N_g=8$ for QCD, and both equal to one for QED.
$D_T$ and $D_L$ are the spatially transverse and longitudinal
gauge boson propagators at finite temperature $T$ and (electron or quark)
chemical potential $\mu$ obtained by Dyson-resumming
one-loop fermion loops, and $D_{\rm vac}$ is the corresponding
quantity at zero temperature and chemical potential.

Nonanalytic terms in the low-temperature expansion arise from the contribution
\begin{equation}
\label{pressuretransverse}
\frac{P_{T,n_b}}{N_{g}}
=-\int \frac{d^{3}q}{(2\pi )^{3}}\int _{0}^{\infty }\frac{dq_{0}}{\pi }2n_{b}\,\textrm{Im}\ln D^{-1}_T.
\end{equation}
The bosonic distribution function $n_b=1/[\exp(q_0/T)-1]$ restricts the $q_0$
integration to $q_0 \lesssim T$ and $T$ is assumed to be the smallest
mass scale in the problem.
Consistently dropping contributions proportional to $T^4$ in the pressure
($T^3$ in the entropy), which for
$T/\mu \ll g$ are beyond our
perturbative accuracy, it turns out that we only need the
$T\to0$ limit of the inverse propagator $D_T^{-1}$, and
only the lowest orders in $q_0/q$ and $q_0/\mu$:
\begin{eqnarray}
\re \,D_{T}^{-1} & = & q^{2}\left(1+O(\g^2)\right) \label{approxReDT} \nonumber\\
 &+&\!\! \left( \frac{\geff ^{2}\mu ^{2}}{\pi ^{2}q^{2}}-1+O(\g^2 q^0)+
O(\g^2 q^2/\mu^2)\right) q_{0}^{2}\nonumber \\
 &+&\!\! O(\g^2 q_{0}^{4}),
\end{eqnarray}
\be
\im \,D_{T}^{-1} = -\frac{\geff ^{2}q_{0}}{48\pi q^{3}}\left( q^{2}-q_{0}^{2}\right) \left( 12\mu ^{2}+3q^{2}+q_{0}^{2}\right) \theta (2\mu -q)\label{approxImDT} 
\ee
where $\geff^2=g^2N_f$ for QED and $g^2N_f/2$ for QCD.

Keeping only the leading terms in the limit $q_0\to0$ gives
\be
\label{ImlnPiTapprox}
\textrm{Im}\ln D_T^{-1}
\simeq \arctan \frac{-\geff ^{2}(4\mu ^{2}+q^2)q_{0}\theta (2\mu - q)}{16 \pi q^{3}}.
\ee
Inserting this approximation into Eq.~(\ref{pressuretransverse}) 
leads to the integral
\bea
\label{integralimprovement1}
&&\int _{0}^{2\mu}dq\, q^{2}\arctan \frac{q_{0}(4\mu ^{2}+q^{2})}{q^{3}}\nonumber\\
&&\simeq \frac{4\mu ^{2}}{3}q_{0}\left( \ln \frac{2\mu}{q_{0}}+\frac{5}{2}\right) +O(q^{5/3}_{0}).
\eea
Performing the $q_0$ integration then gives the following 
contribution to the entropy
$S=\6P/\6T$ (per unit volume):
\bea
\label{nonfermientropyT}
\frac{S_{T,n_b}}{N_{g}}&=&\frac{\geff ^{2}\mu ^{2}T}{36\pi ^{2}}\left( \ln \frac{32\pi \mu }{\geff ^{2}T}+1+\gamma _{E}-\frac{6}{\pi ^{2}}\zeta '(2)\right)\nonumber\\&& +O(T^{5/3}).
\eea

While this reproduces the coefficient of the anomalous $T\ln T^{-1}$
term reported in \cite{Chakravarty:1995}, the coefficient under
the logarithm as well as the suppressed $O(T^{5/3})$-contribution
are still incomplete.

To complete the term linear in $T$, one has to perform an
exactly analogous calculation of the longitudinal contribution,
which involves
\be
\textrm{Im}\ln D_L^{-1} \simeq \frac{\geff ^{2}(4\mu ^{2}-q^2)q_{0}
\theta(2\mu -q)/(8\pi q)}{q^{2}+(\geff ^{2}\mu ^{2})/\pi ^{2}}\,,
\ee
and when inserted into (\ref{pressurecomplete}) contributes
\be\label{entropyL}
\frac{S_{L,n_b}}{N_{g}}= \frac{\geff ^{2}\mu ^{2}T}{24\pi ^{2}}
\left(\ln \frac{\geff ^{2}}{4\pi ^{2}}+1\right)+O(\geff ^{4})+O(T^3).
\ee

Finally, the remaining parts of (\ref{pressurecomplete}), which
do not involve the bosonic distribution function, yield
\be
{S_{non-n_b}\0N_g} = -{\g^2\08\pi^2}\mu^2 T.
\ee

The latter contribution matches exactly the one 
from the standard perturbative result 
to order $g^2$ \cite{Kap:FTFT}, while
the contributions (\ref{nonfermientropyT}) and (\ref{entropyL})
depend on having
$T/\mu \ll g$. In this region, all of the contributions
listed so far are negligible compared to
the zero-temperature contribution $\sim g^4 \mu^4$ in the pressure 
(which is only
partially included in (\ref{pressurecomplete})). However, by considering
instead the entropy (and further below the specific heat),
the above contributions become the dominant ones.

Adding them all together, we obtain
\bea\label{Sseries}
{S-S_0\0N_g}&=& \frac{\geff ^{2}\mu^2 T}{36\pi ^{2}}
\left( \ln{4\g\mu\0\pi^2T}-2+\gamma_{E}-\frac{6}{\pi ^{2}}\zeta '(2)
\right)\nonumber\\&& 
+\, c_{5/3} T^{5/3} + c_{7/3} T^{7/3} + O(T^3)\,,
\eea
where $S_0$ is the ideal-gas value of the entropy per unit volume.

To also obtain completely the coefficients of the
terms in the low-temperature expansion
which involve fractional powers of $T$
we need to include more terms of 
(\ref{approxReDT}) and (\ref{approxImDT})
than those kept in (\ref{integralimprovement1}). 
A lengthy calculation, whose details will be discussed elsewhere, gives
\bea
\label{c53}
c_{5/3} &=& -\frac{8\;2^{2/3}\Gamma (\frac{8}{3})\zeta (\frac{8}{3})}{9\sqrt{3}\pi ^{11/3}}(\geff \mu )^{4/3}\,,\\
\label{c73}
c_{7/3} &=& \frac{80\;2^{1/3}\Gamma (\frac{10}{3})\zeta (\frac{10}{3})}{27\sqrt{3}\pi ^{13/3}}(\geff \mu )^{2/3}\,.
\eea
Setting $T/\mu\sim \g^{1+\delta}$ with $\delta>0$, one finds that
the terms in the expansion (\ref{Sseries}) correspond to
the orders $\g^{3+\delta}\ln(c/\g)$, $\g^{3+(5/3)\delta}$,
and $\g^{3+(7/3)\delta}$, respectively, with a truncation error of
the order $\g^{3+3\delta}$.
Hence, the expansion parameter in this low-temperature
series is $T/(\g\mu)$, which is also the scaleless parameter appearing
in the argument of the leading logarithm (remarkably however only after
the transverse and the longitudinal contributions have been added
together). 
The combination $\g\mu$ is the scale of the Debye mass
at high chemical potential, whose leading-order value is $m_D=\g \mu/\pi$.
In fact, the calculation of the coefficients in (\ref{Sseries})
required keeping the leading-order ``hard-dense-loop'' (HDL)
part of the gauge boson propagator
\cite{Braaten:1990mz,Altherr:1992mf
}, 
in particular the
dynamic screening in (\ref{approxImDT}), but also
a HDL correction to the real part of the transverse
self energy in (\ref{approxReDT}). The above calculation is
therefore in a certain sense another application
of HDL resummation \cite{Altherr:1992mf},
which thus turns out to be necessary also for a perturbative
treatment of the low-temperature regime $T/\mu \ll g$.

\begin{figure}
\vspace{-1cm}
\includegraphics[width=0.64%
\linewidth]{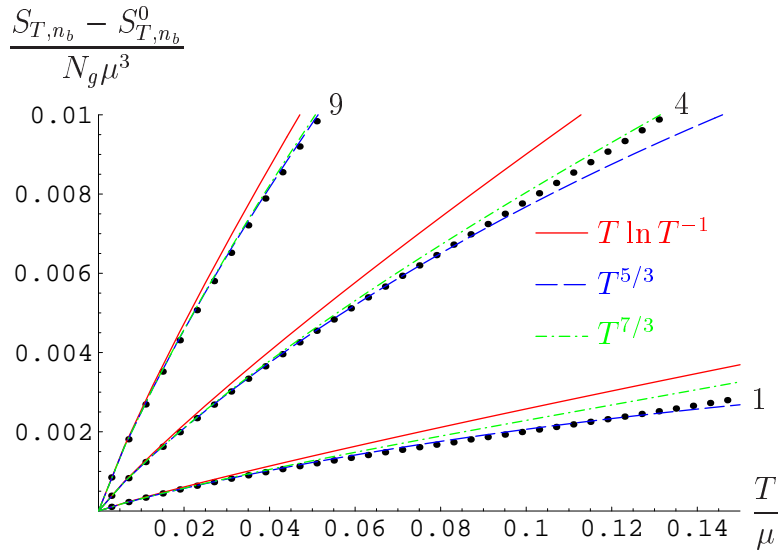}
\caption{Transverse $n_b$-contribution to the interaction part 
of the low-temperature entropy density
in the large-$N_f$ limit for the three values
$\g^2=1,4,9$. The heavy dots give the exact numerical results;
the full, dashed, and dash-dotted lines correspond to our
perturbative result up to and including the $T\ln T^{-1}$, $T^{5/3}$,
and $T^{7/3}$ contributions.\label{figST}}
\end{figure}

As a check on our result and also as a test of its convergence
properties, we compare the anomalous transverse contributions $S_{T,n_b}$ with
those of the exactly (albeit only numerically)
solvable large-$N_f$ limit \cite{Ipp:2003jy} in Fig.~\ref{figST}.
We find good convergence to the exact result
as long as $T/\mu \lesssim \g/(2\pi^2)$.
This is also the region where the complete large-$N_f$ result for
the low-temperature entropy \cite{Ipp:2003jy} has the anomalous property of
exceeding the ideal-gas value.

Our results do not, however, seem to agree with
the results of Ref.~\cite{Boyanovsky:2000bc} which recently questioned the
presence of a term $\propto \alpha T\ln T^{-1}$. The (renormalization group
resummed) result
reported therein rather corresponds to a leading nonanalytic
$\alpha T^3 \ln T$ term when expanded out perturbatively,
which is in fact the type of nonanalytic terms that
appear already in regular Fermi-liquids \cite{Carneiro:1975}.

For potential phenomenological applications in astrophysical
systems, the specific
heat $C_v$ at constant volume and number density is of
more direct interest than the entropy density that we have calculated so far.
The former (per unit volume) is given by \cite{LL:V-Cv}
\be
C_{v}=T\left\{ \left( \frac{\partial S}{\partial T}\right) _{\mu }-{\left( \frac{\partial \mathcal N}{\partial T}\right) ^{2}_{\mu }}{\left( \frac{\partial \mathcal N}{\partial \mu }\right)^{-1} _{T}}\right\},
\ee
where $\mathcal N$ is the number density,
but to the order of accuracy of our expansions, $C_v$ can be
simply obtained as the logarithmic derivative of the entropy:
\be
C_{v}=T\left( \frac{\partial S}{\partial T}\right) _{\mu }+
O(T^3)\,.
\ee

Explicitly, the result is
\begin{eqnarray}
\frac{C_{v}-C_v^0}{N_{g}} & = & \frac{\geff ^{2}\mu ^{2}T}{36\pi ^{2}}\left( \ln \frac{4\geff \mu }{\pi^{2}T}-3+\gamma _{E}-\frac{6}{\pi ^{2}}\zeta '(2)\right) \nonumber\\
&+&\!
{5\03} c_{5/3} T^{5/3} + {7\03} c_{7/3} T^{7/3}
+O(
T^3). 
\label{specificheatNLO}
\end{eqnarray}
with $C_v^0=NN_f \mu^2 T/3 + O(T^3)$, and $c_{5/3}$, $c_{7/3}$ given
by Eqs.~(\ref{c53}), (\ref{c73}).

\begin{figure}[t]
\includegraphics[width=0.64%
\linewidth]{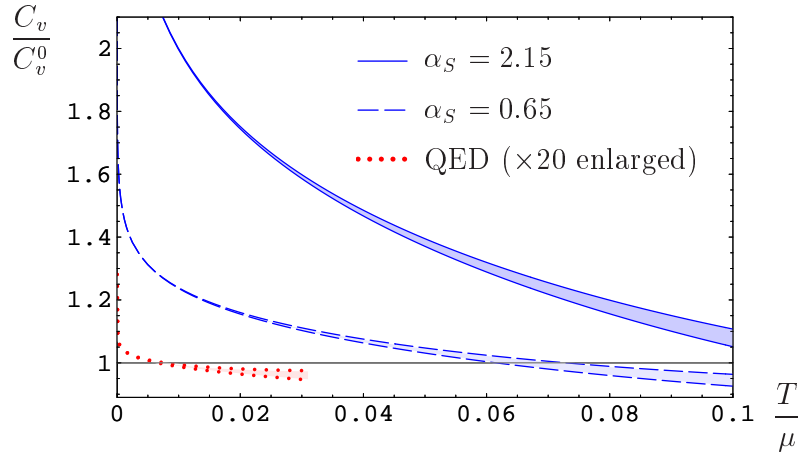}
\caption{The perturbative result for the specific heat,
normalized to the ideal-gas value,
to order $T^{5/3}$ and $T^{7/3}$ (lower and upper curves, respectively)
for two particular values of $\alpha_s$ in two-flavor QCD (chosen
for comparability to Ref.~\cite{Boyanovsky:2000bc}) and
$\g\approx 0.303$ for QED. The deviation of the QED result from
the ideal-gas value is enlarged by a factor of 20, and the
plot terminates where the expansion parameter $(\pi^2 T)/(\g\mu)\approx 1$.\label{figspecificheat}}
\end{figure}

For illustrative purposes,
we evaluate the ratio of $C_v$ as given by (\ref{specificheatNLO}) to the
ideal-gas value $C_v^0$ for QCD with two massless quark flavors in Fig.~\ref{figspecificheat},
using alternatively two values for $\alpha_s$ which have
been used also in Ref.~\cite{Boyanovsky:2000bc} and which
correspond to one-loop running couplings with renormalization
point $0.5$ GeV (full line) and 1 GeV (dashed line).
The shaded bands shown are limited from below and above 
by the results to order
$T^{5/3}$ and $T^{7/3}$, respectively, and thus indicate the
quality of the low-temperature expansion.
One may interpret these results as roughly corresponding to
QCD with a quark chemical potential of $0.5$ GeV and the total variation
corresponding to different renormalization schemes with
minimal subtraction scale varied between $\mu$ and $2\mu$.
The critical temperature for the color superconducting phase
transition may be anywhere between 6 and 60 MeV \cite{Rischke:2003mt},
so the range $T/\mu \ge 0.012$ in Fig.~\ref{figspecificheat} 
might correspond to normal quark matter. 
While it is certainly questionable to apply perturbative results
for $\alpha_s \gtrsim 0.65$, Fig.~\ref{figspecificheat}
suggests that the anomalous feature
of an excess of the specific heat over its ideal-gas value
may possibly
come into play in astrophysical situations,
in particular in the cooling of (proto-)neutron stars 
\cite{Iwamoto:1980eb,Carter:2000xf
}.
This should be contrasted with the
ordinary perturbative estimate for $C_v/C_v^0$ 
based on the well-known \cite{Kap:FTFT}
exchange term $\propto g^2$ (which, as we have shown, requires 
$T/\mu \gg g$). The latter would
predict $C_v/C_v^0\lesssim 0.6$ for $\alpha_s \gtrsim 0.65$.

For completeness, we also give the numerical results corresponding
to QED, where $\g\approx 0.303$. Here the range of temperature,
where the specific heat exceeds the ideal-gas value, and
the deviations from the latter, are much smaller (the deviations
from the ideal-gas value have been enlarged by a factor of 20
in Fig.~\ref{figspecificheat} to make them more visible).

To summarize, we have presented 
a quantitative
evaluation of the leading contributions to the entropy and specific
heat of high-density QCD and QED in the regime $T/\mu \ll g \ll 1$, which
is dominated by non-Fermi-liquid behavior. While the effect remains
small in QED, it seems conceivable that the anomalous terms in
the specific heat play a noticeable role in the thermodynamics of
a normal quark matter component
of neutron or proto-neutron stars.

\acknowledgments

A.~Ipp and A.~Gerhold have 
been supported by the Austrian Science Foundation FWF,
project no. 14632-PHY and 16387-N08, respectively.


\end{document}